\documentclass{acm_proc_article-sp}
\usepackage[T1]{fontenc}
\usepackage[latin9]{inputenc}
\usepackage{amssymb}
\usepackage{amstext}
\usepackage{graphicx}
\usepackage{subscript}
\usepackage[all]{xy}
\makeatletter

\providecommand{\tabularnewline}{\\}

\@ifundefined{showcaptionsetup}{}{%
 \PassOptionsToPackage{caption=false}{subfig}}
\usepackage{subfig}
\makeatother

\begin{document}

\title{Quantitative Analysis of Active Cyber Defenses Based on Temporal Platform Diversity\titlenote{This work is sponsored by the Department of Defense under Air Force Contract FA8721-05-C-0002. Opinions, interpretations, conclusions and recommendations are those of the author and are not necessarily endorsed by the United States Government.}}

\numberofauthors{3} 

\author{
\alignauthor
Kevin M. Carter\\
       \affaddr{MIT Lincoln Laboratory}\\
       \affaddr{244 Wood St.}\\
       \affaddr{Lexington, MA 02420}\\
       \email{kevin.carter@ll.mit.edu}
\alignauthor
Hamed Okhravi\\
       \affaddr{MIT Lincoln Laboratory}\\
       \affaddr{244 Wood St.}\\
       \affaddr{Lexington, MA 02420}\\
       \email{hamed.okhravi@ll.mit.edu}
\alignauthor James Riordan\\
       \affaddr{MIT Lincoln Laboratory}\\
       \affaddr{244 Wood St.}\\
       \affaddr{Lexington, MA 02420}\\
       \email{james.riordan@ll.mit.edu}
}

\maketitle
\begin{abstract}
Active cyber defenses based on temporal platform diversity have been proposed 
as way to make systems more resistant to attacks. These defenses change the properties of the 
platforms in order to make attacks more complicated. Unfortunately, little work has been done 
on measuring the effectiveness of these defenses. In this work, we use four different approaches to 
quantitatively analyze these defenses; an abstract analysis studies the algebraic models of 
a temporal platform diversity system; a set of experiments on a test bed
measures the metrics of interest for the system; a game theoretic analysis 
studies the impact of preferential selection of platforms and derives an optimal strategy; 
finally, a set of simulations evaluates the metrics of interest on the models. Our results from 
these approaches all agree and yet are counter-intuitive. We show that although platform diversity can mitigate some attacks, it can be detrimental for others. We also illustrate that the benefit from these systems 
heavily depends on their threat model and that the preferential selection of platforms can achieve better protection.

\end{abstract}




\section{Introduction}
Developing secure systems is difficult and costly. The high cost of effectively mitigating all vulnerabilities and the
far lesser cost of exploiting a single one creates an environment which advantages cyber attackers. New cyber defense
paradigms have been proposed to re-balance the landscape and create uncertainty for the attackers \cite{nitrd}. One such
paradigm is active defense based on temporal platform diversity.

Temporal platform diversity (or simply platform diversity) techniques dynamically change the properties of a computing
platform in order to complicate attacks. Platform properties refer to hardware and operating system (OS) attributes such
as instruction set architecture (ISA), stack direction, calling convention, kernel version, OS distribution, and machine
instance. Various platform diversity techniques have been proposed in the literature. Emulation-based techniques change
the calling sequence and instruction set presented to an application \cite{genesis}; multivariant execution techniques
change properties such as stack direction or machine description using compiler generated diversity and virtualization
\cite{multivariant,multivariant2,multivariant3,machinedesc}; migration-based techniques change the hardware and
operating system of an application using containers and compiler-based checkpointing \cite{talent}; server
diversification techniques rotate a server across multiple platforms and software stacks using network proxies
\cite{webserver}; self cleansing techniques change the machine instance by continuously rotating across many virtual
machines and re-imaging the inactive ones \cite{scit,scit2,scit3}. The key point to these techniques is that platform
diversity does not remove vulnerabilities, it just makes them more complicated to exploit.

Unfortunately little work has been done on understanding and quantifying the impact of temporal platform diversity on the security of a system. The following important questions often remained unanswered: how often should the platform change? How diverse are various platforms? What is the quantitative impact of platform diversity on attacks? What are the important parameters and factors for the effectiveness of a platform diversity system? Should the next platform or platform configuration be selected uniformly or preferentially? What is the optimal strategy for controlling a platform diversity system?

In this work, we take various approaches to quantitatively analyze and study active cyber defenses based on platform diversity. We use a migration-based technique (i.e.~Talent \cite{talent}) in our study and quantitatively measure its properties using four approaches. First, we perform an abstract analysis of platform diversity using algebraic models, combinatorics, and Markov models. Second, we implement the platform diversity system on a testbed and perform experiments with it using real exploits to quantitatively measure its properties. Third, we analyze the temporal migration patterns using game theory and derive an optimal strategy for selecting the next platform for five operating systems: CentOS, FreeBSD, Debian, Fedora, and Gentoo. Fourth, we model the platform diversity system running the above operating systems with and without the optimal strategy and simulate its impact on an attacker trying to compromise the system.

The quantitative results from our four approaches are consistent and in some cases surprising and counter-intuitive. Our findings indicate that although platform diversity can help protect a system against some attacks (specifically, persistent service disruption attacks), it can negatively impact security for others (specifically, fast, single compromise attacks). We show that the benefit of platform diversity techniques heavily depends on their threat models and that operational requirements and temporal parameters must be understood and fine tuned for these techniques to be effective. Moreover, we prove that preferential selection of platforms can achieve better protection and derive the optimal strategy for determining temporal patterns (i.e.~the next platform).

The rest of the paper is organized as follows. Section \ref{PDB} provides a brief overview of platform diversity techniques and Talent. Section \ref{TM} describes the threat model used throughout the paper. Section \ref{AA} discusses our abstract analysis approach and its results. Section \ref{Exper} discusses our testbed experiments and measurements performed on a real system. Section \ref{GTOC} provides the game theoretic analysis of temporal patterns and the optimal solution. Section \ref{Sim} explains our simulation approach and results. Section \ref{LLD} enumerates a number of lessons learned and discusses our findings. We discuss the related work in Section \ref{RW} before concluding the paper in Section \ref{Conc}.

\section{Platform Diversity Background}
\label{PDB}

In this section, we briefly describe the active defense techniques based on platform diversity. We provide enough background for understanding the rest of the paper. More details about each technique can be found in its original publication.

Temporal platform diversity techniques change platform properties in order to make attacks more complicated. They often rely on temporal changes (e.g.~VM rotation), diversity (e.g.~multivariant execution), or both (e.g.~migration-based techniques) to protect a system. These techniques are often implemented using machine-level or operating system-level virtualization, compiler-based code diversification, emulation layers, checkpoint/restore techniques, or a combination thereof. Emulation-based techniques such as Genesis \cite{genesis} often use an application-level virtual machines such as Strata \cite{strata} or Valgrind \cite{valgrind} to implement instruction set diversity. In some cases, multiple instances are executed and a monitor compares their results. Multivariant execution techniques such as Reverse stack \cite{reversestack} (also called N-variant systems \cite{nvariant}) use compiler-based 
techniques to create diverse application code by replacing sets of instructions with semantically equivalent ones. Migration-based techniques such as Talent \cite{talent} use operating system-level virtualization (containers) to move an application across diverse architectures and operating systems. Platform diversity can also be achieved at a higher abstraction level by switching between different implementations of servers \cite{webserver}. These techniques either do not preserve the state (e.g.~a web server) or they preserve it using high level configuration files (e.g.~DNS server). Finally, self-cleansing techniques such as SCIT \cite{scit} only change the current instance of the platform without diversifying it. The main goal, in this case, is bringing the platform to its pristine state and removing persistence of attacks.

In this work, we use a migration-based platform diversity technique, Talent, in our analyses. Since Talent provides both temporal changes (periodic migrations) and diversity (multiple operating systems), it is a good candidate for capturing the major effects and dynamics in our quantitative analysis.

\subsection{Talent}
\underline{T}rusted dyn\underline{a}mic \underline{l}ogical h\underline{e}teroge\underline{n}eity sys\underline{t}em (Talent) \cite{talent} is a framework for live-migrating critical applications across heterogeneous platforms. Talent has several design goals: i) heterogeneity at the instruction set architecture level, ii) heterogeneity at the operating system level,
iii) preservation of the state of the application, including the execution state, open files and sockets, and iv) working with a general-purpose, system language such as C. Talent uses two key concepts, operating-system-level virtualization and portable checkpoint compilation, to address the challenges involved in using heterogeneous platforms, including binary incompatibility and the loss of state and environment.

\subsubsection{Environment Migration}

The environment of an application refers to the filesystem, configuration files, open files, network connections, and open devices. Many of the environment parameters can be preserved using virtual machine migration. However, virtual machine migration can only be accomplished using a homogeneous operating system and hardware. 

Talent uses operating-system-level virtualization to sandbox an application and migrate the environment. In operating-system-level virtualization, the kernel allows for multiple isolated user-level instances. Each instance is called a container (jail or virtual environment). This type of virtualization can be thought of as an extended \texttt{chroot} in which all resources (devices, filesystem, memory, sockets, etc.) are virtualized. Hence, the semantic information that is often lost in hardware virtualization is readily available in operating-system-level virtualization.

Talent is implemented using the OpenVZ \cite{openvz}. When migration is requested, Talent migrates the container of the application from the source machine to the destination machine. This is done by synchronizing the filesystem of the containers, migrating the IP address of the source container's virtual network interface, and transferring the state of each TCP socket (\texttt{sk\_buff} structure of the kernel). IPC and signal migration is supported in Talent through the underlying virtualization layer.

\subsubsection{Process Migration}

Migrating the environment is only one step in backing up the system because the state of running programs must also be migrated. To do this, a method to checkpoint running applications must be implemented. 

Talent uses a portable checkpoint compiler (CPPC \cite{cppc}) to migrate a process state. CPPC allows Talent to checkpoint the state of the binary on one architecture and recover the state on a different architecture. To perform checkpointing, the code is precompiled to find restart-relevant variables. These variables and their memory locations are then registered in the
checkpointing tool. When checkpointing, the process is paused and the values of the memory locations are dumped into a file. The checkpointing operation must occur at safe points in the code to generate a consistent view. At restart, the memory of the destination process is populated with the desired variable values from the checkpoint file. Some portions of the code are re-executed in order to construct the entire state. The checkpoint file itself must have a portable format to achieve portability across 32-bit/64-bit architectures and little/big endian machines. Talent uses the HDF5 portable file format \cite{HDF5} for its checkpoints.

Talent has been implemented on Intel Xeon 32-bit, Intel Core 2 Quad 64-bit, and AMD Opteron 64-bit processors. It has also been tested with Gentoo, Fedora (9, 10, 11, 12, and 17), CentOS (4, 5, and 6.3), Debian (4, 5, and 6), Ubuntu (8 and 9), SUSE (10 and 11), and FreeBSD 9 operating systems.

\section{Threat Model}
\label{TM}

We discuss multiple threat models in this paper but analysis shows that they share common features. To make the analysis
more precise, we explicitly describe the core threat model in this section. Variations upon the core threat model are
described in the other sections as appropriate.

In our model, the defender has a number of different platforms to run a critical application. The attacker has a set of exploits (attacks) that are applicable against some of these platforms, but not the others. We call the platforms for which the attacker has an exploit ``vulnerable'' and the others ``invulnerable.'' In a strict systems security terminology, vulnerable does not imply exploitable; without loss of generality, we only consider exploitable vulnerabilities. An alternative interpretation of this threat model is that the vulnerabilities are exploitable on some platforms, but not on the other ones.

The defender does not know which platforms are vulnerable and which are invulnerable, nor does he have detection capabilities for the deployed exploits. This scenario, for example, describes the use of zero-day exploits by attackers, for which no detection mechanism exists by definition.

Since there is little attempt to isolate the inactive platforms in platform diversity systems, we assume that all platforms are accessible by the attacker, and the attacker attempts to exploit each one. Moreover, we assume that the defender does not have a recovery or re-imaging capability to restore a compromised platform. The reason for this assumption is two fold: first, typical recovery methods (e.g.~clean installation of the operating system) can take a long time to complete and unless the defender has many spare platforms, it is hard to accomplish it effectively. A large number of spare platforms also implies large hardware cost for these systems (e.g.~SCIT \cite{scit}). Second, and more importantly, the goal of this paper is to study the impact of temporal platform changes, not the effectiveness of recovery capabilities being used. Hence, once exploited, a platform remains vulnerable and the attacker requires no staging time in the future.

The attacker's goal is what creates the variations in our threat model. For example, one success criteria may be for the adversary to compromise the system for a given period of time to cause irreversible damage (e.g.~crash a satellite), while a different success criteria gives the attacker gradual gain the longer the system is compromised (e.g.~exfiltration of information). These variations and their impact on the effectiveness of temporal platform diversity are discussed in the subsequent sections.

\section{Abstract Analysis}
\label{AA}

Much of the analysis of one system using temporal platform diversity
as a security strategy applies to any such system. In this section
we will specify an abstraction that generically describes temporal
platform diversity, which is useful as a security strategy in protecting against attacks that have a temporal requirement. We analyze a number of cases where this holds. To better convey the meanings, 
we use simple phrases which are examples of these cases to refer to them: 
crash the satellite, sneaking past a sensor, and  exfiltration over a difficult channel.

\begin{enumerate}
\item Crash the Satellite - once the attacker has controlled the system
continuously for a specified time period, the game is over and the
attacker has won.
\item Sneaking past a sensor - the attacker needs control for a specified
time period for an attack to succeed. Once the attacker has control
for the period, the attacker's payoff function starts to increase from
zero. The scenario can either be continuous or finite duration.
\item Exfiltration over a difficult channel - the attacker needs control
for a specified time period for an attack to succeed. Once the attacker
has control for the period, the attackers payoff function starts to
increase from a positive value.
\end{enumerate}
We can categorize the problem space according to a number of properties:
\begin{itemize}
\item the attackers temporal requirement can either be aggregate or continuous,
\item the attackers payoff can be either fractional or binary (all or nothing),
\item the period of usage (equivalently attack window) can either be of fixed
duration or ongoing, and
\item the migration models include random with repeat, random without repeat,
and periodic permutation
\end{itemize}
We will define the abstract model of a temporal platform diversity
system ${\cal S}$ as a system that migrates through a finite fixed collection
of states $\left\{ s_{i}\right\}$. Each state either has or does
not have a property exploitable by the attacker which we call vulnerable. In the first approximation to the model we assume that the states are fully independent. We will use the notation presented in Table \ref{table:notation}.

\begin{table}[t]
\begin{tabular}{|l|l|}
\hline 
$m$  & number of vulnerable states\tabularnewline
$n$  & number of invulnerable states \tabularnewline
$s^{k}$ & state at migration step $k$\tabularnewline
$v\left(s^{k}\right)$ & state at migration step $k$ is vulnerable\tabularnewline
$\neg v\left(s^{k}\right)$ & state at migration step $k$ is not vulnerable\tabularnewline
$P\left(v\left(s^{k}\right)\right)$ & probability that $v\left(s^{k}\right)$\tabularnewline
$P_{vv}$ & $P\left(v\left(s^{k+1}\right)|v\left(s^{k}\right)\right)$ \tabularnewline
$P_{ii}$ & $P\left(\neg v\left(s^{k+1}\right)|\neg v\left(s^{k}\right)\right)$\tabularnewline
\hline 
\end{tabular}
\caption{Notation describing platform diversity system}
\label{table:notation}
\end{table}

\subsection{Attacker Aggregate Control}

When the attacker requires only aggregate control, there are two main subcategories according to the
attacker's payoff.  The fractional case is trivially determined by the ratio of $m$ and $n$. In the binary case, wherein
the attacker wins by controlling a specified fraction of the vulnerable time, the defender may optimize via an initial
subselection of states in a process reminiscent of gerrymandering. For example, if $m=3$ and $n=2$ and the attacker wants
to control greater than 50\% of the time, then the defender should simply expect to lose should all platforms be
utilized. By contrast if the defender randomly subselects two then the defender can reduce the attacker's expectation of
winning to
\[
\frac{C\left(3,2\right)}{C\left(5,2\right)}=\frac{3}{10}=30\%,
\]
where $C\left(x,y\right)=\frac{x!}{y!(x-y)!}$ is the combinatorial choice function. Here the value of 2 as the number of
platforms chosen.

Generally, if $p$ is the percentage of time that the attacker requires for success and we subselect $j$ platforms from the total $m+n$,
then the probability of attacker success is
\[
P_{success}=\sum_{i=\lceil p\cdot j\rceil}^{\min\left(m,\, j\right)}\frac{C\left(m,\, i\right)\cdot C\left(n,\, j-i\right)}{C\left(m+n,\, j\right)},
\]
in the steady-state model.

\subsection{Attacker Continuous Control}

When the attacker requires continuous control, the defender can use the subselection strategy as above as well as
leveraging conditional probabilities. These conditional probabilities are given in Table
\ref{TableConditionalProbabilities}.

\begin{table*}[t]
\centering{}%
\begin{tabular}{|c|c|c|c|c|c|}
\hline 
Repeat & {\footnotesize $\!\!$Vuln$\!\!$} & {\footnotesize $\!\!$$\neg$Vuln$\!\!$} & {\footnotesize $\!\! P\left(v\left(S_{i+1}\right)\right)\!\!$} & {\footnotesize $\!\! P\left(v\left(s_{i+1}\right)\!\mid\! v\left(s^{k}\right)\right)\!\!$} & {\footnotesize $\!\! P\left(v\left(s_{i+j+1}\right)\!\mid\! v\left(s_{i+j}\right)\&\ldots\&v\left(s^{k}\right)\right)\!\!$}\tabularnewline
\hline 
\hline 
Without & $m$ & $n$ & $\frac{m}{m+n}$ & $\frac{m-1}{m+n-1}$ & $\frac{m-j}{m+n-j}$\tabularnewline
\hline 
With & $m$ & $n$ & $\frac{m}{m+n}$ & $\frac{m}{m+n}$ & $\frac{m}{m+n}$\tabularnewline
\hline 
\end{tabular}\caption{Conditional Probabilities}
\label{TableConditionalProbabilities}
\end{table*}
Here, we observe that $\frac{m}{m+n}>\frac{m-j}{m+n-j}$ so long as
$n$ and $j$ are both greater than zero. As such, migrating \emph{without} immediate
repeat, while not influencing the fraction of vulnerable platforms
selected, tends to reduce successful sequences. We note that the influence is
greater when a smaller number of platforms is used. Our later experiment will
use 3 vulnerable and 2 invulnerable platforms which is a sufficiently
small number to have a strong influence upon the conditional probabilities.

This reduces to the Markov chain
\[
\xymatrix{ &  & V\ar@(ru,rd)[]^{P_{vv}}\ar@/^{1pc}/[dd]^{1-P_{vv}}\\
*+[F-]{\text{start}}\ar[rru]^{P_{v}}\ar[rrd]_{P^{k}=1-P_{v}}\\
 &  & I\ar@(ru,rd)[]^{P_{ii}}\ar@/^{1pc}/[uu]^{1-P_{ii}}
}
\]
which can be used to confirm that the steady state

\subsection{Attacker Fractional Payoff Model}

The steady state of attacker control of the system can be modeled
using Markov chains with states $I$ and $V$ referring to invulnerable and vulnerable respectively. While the simple Markov model describing the
transitions $\left\{ I,V\right\} \longrightarrow\left\{ I,V\right\}$ describes
the base behavior of the system, it does not naturally capture the
notion of repeated vulnerable states. We can adapt this chain to one
with a richer collection of states 
{\small
\[
\left\{ I,IV,IV^{2},\ldots,IV^{n-1},V^{n}\right\} \longrightarrow\left\{ I,IV,IV^{2},\ldots,IV^{n-1},V^{n}\right\} 
\]
}
which support runs of length $n$. The probability of invulnerable
to invulnerable transition is given by 
\[
P_{ii}=P\left(\neg v\left(s_{i+1}\right)\!\mid\!\neg v\left(s^{k}\right)\right)=\frac{n-1}{m+n-1}
\]
 and the probability of vulnerable to vulnerable transition is given
by
\[
P_{vv}=P\left(v\left(s_{i+1}\right)\!\mid\! v\left(s^{k}\right)\right)=\frac{m-1}{m+n-1}
\].
The Markov model looks like

\[
\xymatrix{ & {IV}\ar@/^{1pc}/[ld]_{\footnotesize{1-P_{vv}\!\!}}\ar[rr]^{\footnotesize{P_{vv}}} &  & {IV^{2}}\ar@/^{0.75pc}/@{-}[llld]_{\footnotesize{1-P_{vv}}}\ar@{.>}[dr]\\
{I}\ar@/^{1pc}/[ru]^{\footnotesize{1-P_{\neg v}}}\ar@(l,d)[]_{\footnotesize{P_{ii}}} &  &  &  & IV^{n-1}\ar[llll]_{\;\;\;\;\footnotesize{1-P_{vv}}}\ar[ld]^{\footnotesize{P_{vv}}}\\
 &  &  & {V^{n}}\ar@{-}@/_{1pc}/@{-}[lllu]^{\footnotesize{1-P_{v}}}\ar@(l,d)[]_{\footnotesize{P_{vv}}}
}
\]
which has the $(n+1)\times (n+1)$ Markov transition matrix is given by

\[
\left[\begin{array}{ccccccc}
P_{ii} & 1-P_{vv}\\
1-vv_{v} &  & P_{vv}\\
1-P_{vv} &  &  & P_{vv}\\
1-P_{vv} &  &  &  & P_{vv}\\
1-P_{vv} &  &  &  &  & \ddots\\
1-P_{vv} &  &  &  &  &  & P_{vv}\\
1-P_{vv} &  &  &  &  &  & P_{vv}
\end{array}\right]
.\]
This transition matrix has the steady state eigen-vector
{\small
\[
\left[\begin{array}{ccccc}
\frac{n}{m+n} & \; a_{v}\cdot P_{vv} & \; a_{v}\cdot P_{vv}^{2}\;\cdots  & \; a_{v}\cdot P_{vv}^{n-1} & \; a_{v}\cdot\sum_{i=n}^{\infty}P_{vv}^{i}\end{array}\right]
\]
}
where
\[
a_{v}=\frac{m}{m+n}\cdot\left(\frac{1-P_{vv}}{P_{vv}}\right).
\] 

This can be used to compute the steady state behavior of the system.
If the attacker success begins after $n$ steps then the steady state
is given by the right most term in the eigen vector $a_{v}\cdot\sum_{i=n}^{\infty}P_{v}^{i}=\frac{m}{m+n}-a_{v}\cdot\sum_{i=0}^{n-1}P_{v}^{i}$.
If the attacker success includes the steps leading to a run of $n$
steps then we must also include vulnerable states weighted by the
probability that they will become a run of $n$ vulnerable states
and the contribution to the run: the probability that $IV^{n-1}$ will
become $V^{n}$is $P_{V}$, the probability that $IV^{n-2}$ will become
$V^{n}$is $2\cdot P_{V}^{2}$ and so forth. Reducing that equation,
we find that the expected period of attack control $L(n)$ is 
\[
L(n)=1-\frac{\left(1-P_{\neg v}\right)^{-1}+\left(1-P_{v}\right)\sum_{i=0}^{n-1}i\cdot P_{v}^{i-1}}{\left(1-P_{\neg v}\right)^{-1}+\left(1-P_{v}\right)^{-1}}
\]
which is one minus the percentage of time that the defender is in control.

\subsection{Attacker Binary Payoff Model}
\label{ABPM}

In the binary payoff model with random selection (with or without immediate
repeats), the attacker will eventually win so long as it is combinatorially
possible in the same manner that a person flipping a coin will eventually
observe a sequence of ten, or ninety-two, heads in a row. Here metrics
might reasonably be based in the mean time until attacker victory.
These can be analyzed in a fashion similar to the steady state model:

\[
\xymatrix{*+[F-]{I^{\left\{ 1\cdots\infty\right\} }}\ar[dr]^{1-P_{vv}^{n-1}}\ar@/_{6pc}/[dd]^{P_{vv}^{n-1}}\\
*+[F-]{\text{start}}\ar[r]^{\negthickspace\negthickspace\negthickspace\negthickspace\negthickspace\negthickspace\negthickspace\negthickspace\negthickspace\negthickspace\negthickspace\negthickspace\negthickspace\negthickspace P_{*}}\ar[d]_{\footnotesize{P_{v}P_{vv}^{n-1}}}\ar[u]^{1-P_{v}} & *+[F-]{V^{\left\{ 1\cdots n-1\right\} }I^{\left\{ 1\cdots\infty\right\} }}\ar[dl]^{P_{vv}^{(n-1)}}\ar@(ul,ur)[]^{\;\;\;\;\;\;1-P_{vv}^{n-1}}\\
*+[F-]{V^{n}}\ar[d]^{1}\\
*+[F-]{\text{end}}
}
\]
where $P_{*}=P_{v}\left(1-P_{vv}^{n-1}\right)$.  We can use this to evaluate the expected time $L'(n)$ to attack compromise
as the probabilistically weighted sum of all path lengths
\begin{equation}
\begin{split}
L'(n)=&n+\frac{1-P_{v}}{1-P_{ii}}+\\
	&\left(P_{vv}^{1-n}-1\right)\cdot\\
        &\left(\frac{1-n\cdot P_{vv}^{n-1}+\left(n-1\right)\cdot P_{vv}^{n}}{\left(1-P_{vv}^{n-1}\right)\cdot\left(1-P_{vv}\right)}+\frac{1}{1-P_{ii}}\right)
\end{split}
.
\label{eq:time_to_compromise}
\end{equation}
The full derivation can be found in Appendix \ref{derivation}. Hence, in scenarios such as `crash the satellite', Eq.~(\ref{eq:time_to_compromise}) computes the expected time before the adversary is able to take down the service.

\subsection{Finite Duration}
\label{fd}

While the effect of a finite duration scenario with random attacker
starting time does not strictly concern temporal platform diversity systems, the effect
needs to be explained in order to interpret the experimental results.
Let $d$ be the duration of the trial, $a$ be the period that the
attacker must be present.

\begin{figure}[t]
\begin{centering}
\includegraphics[width=0.75\columnwidth]{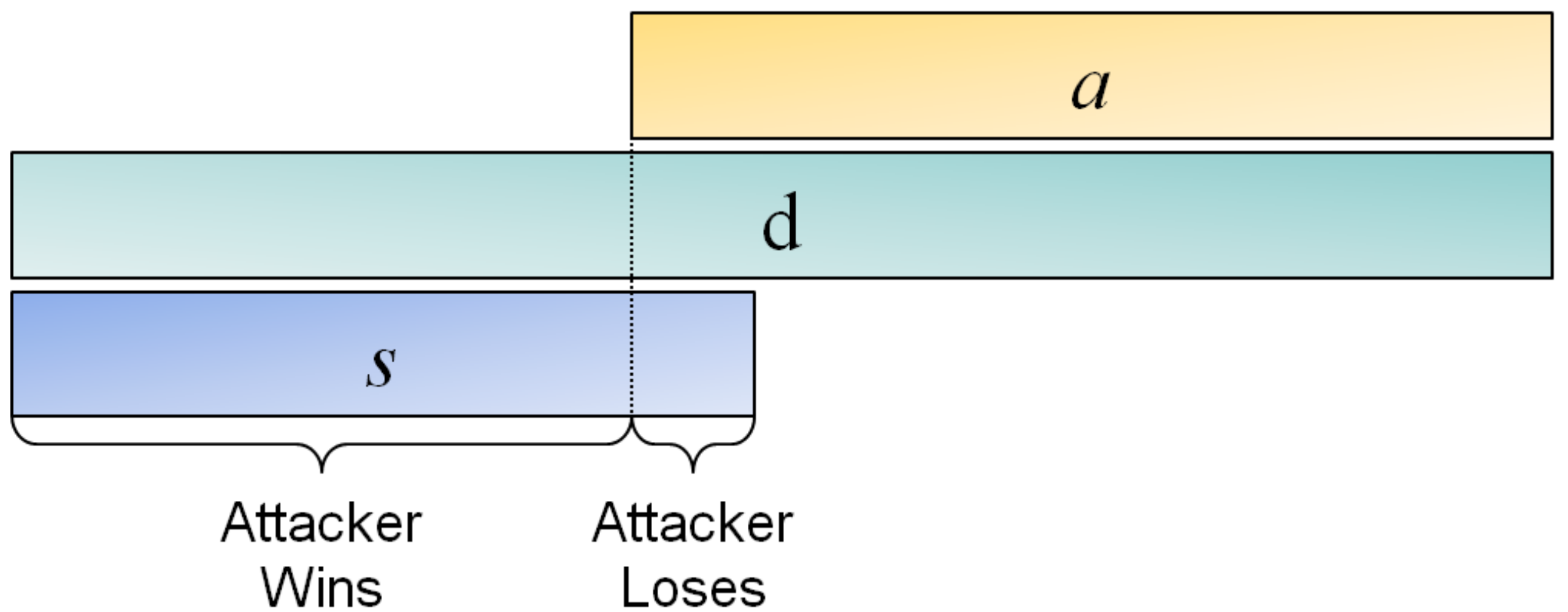}
\par\end{centering}

\caption{Windows Of Opportunity}
\label{window}
\end{figure}

Then the probability that the attack succeeds is given by 

\[
P_{success}=\min\left(1,\max\left(0,\frac{d-a}{s}\right)\right)
\]

As a function of $a$, the variable region is a line of slope $-\frac{a}{s}$.
This is approximated by the graph with one platform where there the
only deterministic influence is this window of opportunity (see Figure
\ref{FigureResults1}).

\begin{figure}[t]
\begin{centering}
\includegraphics[width=0.75\columnwidth]{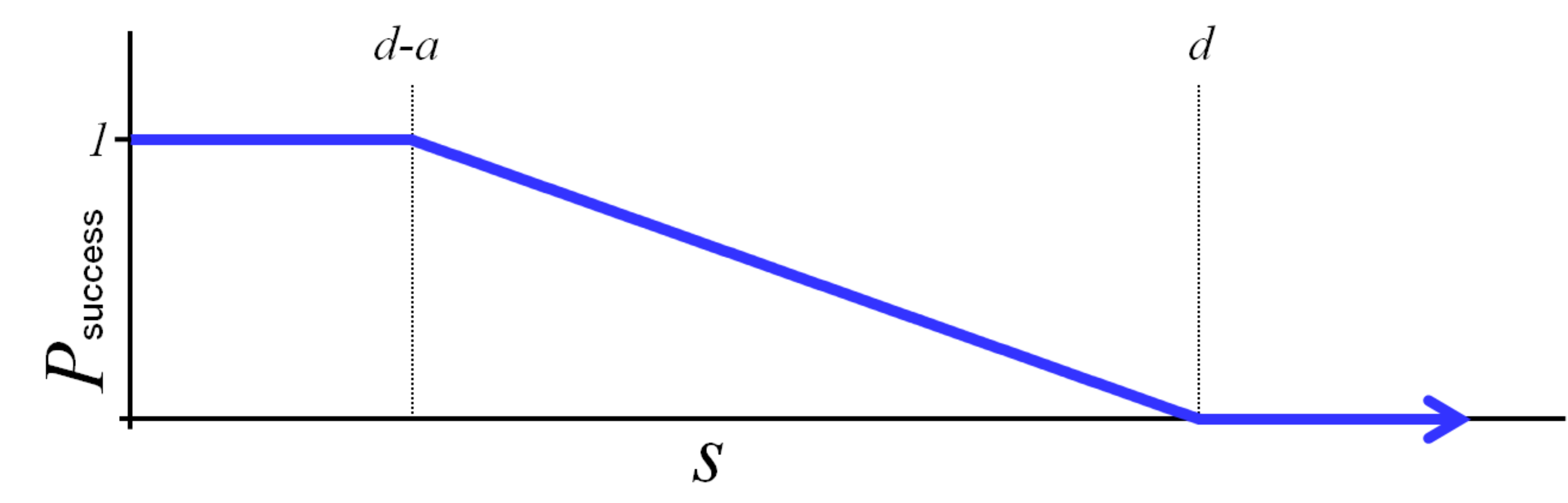}
\par\end{centering}

\caption{Probability}
\label{baseline}
\end{figure}

\subsection{Fractional Effect}
\label{fe}

The final deterministic effect is a sort of fractional effect which occurs when the required attacker duration $a$
passes between various multiples of the duration of each platform. For example, if the duration of the attacker passes
from being a bit less than a single platform duration to a bit more, then the random selection
process means that instead of a single vulnerable platform showing up, two need to show up consecutively. The same thing
happens as we transition from two to three and so on. The result is a sort of downward step in the graphs which is
smoothed but the the initial attack step (e.g.~if there is one vulnerable state but the attacker arrives in its middle,
it might be an inadequate duration for the attacker).

Specific examples of attacks for this form will be given in ...
\begin{enumerate}
\item continuous attacker control / aggregate attacker control
\item periodic permutational / random with repeat / random without repeat
\end{enumerate}
The platform migration model used in the experiment is random without immediate repeat (e.g.~for platforms are $\left\{
A,B,C\right\} $ the the sequence $A\rightarrow B\rightarrow A\rightarrow C$ is permissible while $A\rightarrow
A\rightarrow B\rightarrow C$ is not). 

The fractional effect can be observed in our experiments in Figure \ref{FigureResults1}.

\section{Experiments}
\label{Exper}

\subsection{Experiment Setup}

To perform the experiments, a notional application with C back-end and GUI front-end has been ported to Talent. On the test bed, we have a pool of five different platforms: Fedora on x86, Gentoo on x86, Debian on x86\_64, FreeBSD on x86, and CentOS on x86. The application runs for a random amount of time on a platform before migrating to a different one (i.e.~inter-migration delay). 

The attacker's goal is to control the active platform for some time $T$. Since in a real scenario the vulnerability of the platform is unknown, we may consecutively migrate to multiple vulnerable platforms, in which case the attacker wins. To implement this scenario on the test bed, we launch two real exploits against Talent. The first is the TCP MAXSEG exploit which triggers a divide by zero vulnerability in \texttt{net/ ipv4/tcp.c} (CVE-2010-4165) to perform a DoS attack on the platform. Only the Gentoo platform is vulnerable to this attack. The second attack is the Socket Pairs exploit which triggers a garbage collection vulnerability in \texttt{net/unix/ \\garbage.c} (CVE-2010-4249) to saturates the CPU usage and file descriptors. The Fedora and CentOS platforms are vulnerable to this attack. Our Debian and FreeBSD platforms are not vulnerable to these exploits.

In each configuration, we select $N\in(1,5)$ platforms. For each sample, the application randomly migrates across those $N$ platforms without immediate repeat. In the case of $N = 1$ (baseline), the application remains on the same platform during the entire run. Without loss of generality, the inter-migration delay ($t_{mig}$) is chosen randomly and uniformly from 20-30 seconds. Although we have no reason to believe that these are the appropriate values for a real-world application, we will show later that the actual values
of the inter-migration delay ($t_{mig}$) and attacker's goal ($T$) are inconsequential to our experiments and can be parametrized. 

One or both exploits become available to the attacker at random times during each sample. As a result, zero to three platforms can be breached (zero when the exploit is not effective against the set of platforms and three when both exploits are available and Fedora, CentOS, and Gentoo are in the pool of platforms). When the exploit is launched, its payload reaches all of the platforms in the selected set at once (not one after another). This approach tries to model the behavior of network-based exploits that propagate to all machines within a network very rapidly. Each sample runs for 15 minutes, leading to 300 collected samples for each configuration. We also collect a central log which includes a timestamp, the status of each platform (up or down), and the active platform and a local log (for verification purposes) which also includes finer-grained CPU load for each platform. 

\begin{figure}
\begin{centering}
\includegraphics[width=0.75\columnwidth]{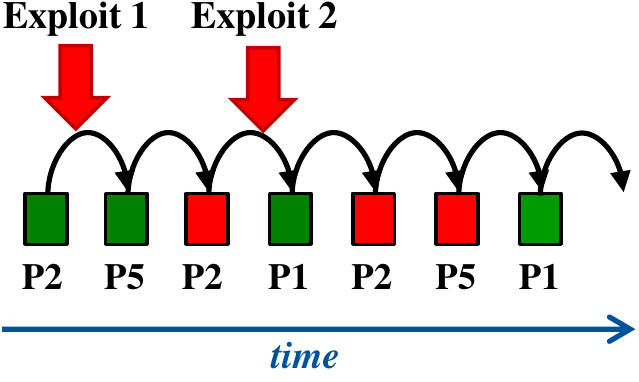}
\caption{A 3-platform sample}
\label{FigureSample}
\end{centering}
\end{figure}

Figure \ref{FigureSample} illustrates one sample with 3 platforms. The red arrows show when exploits are launched. In this case, platforms
2 and 5 are vulnerable to the exploits.

\subsection{Experiment Results}
\label{SS:ExperimentalResults}
\begin{figure}
\begin{centering}
\includegraphics[width=1\columnwidth]{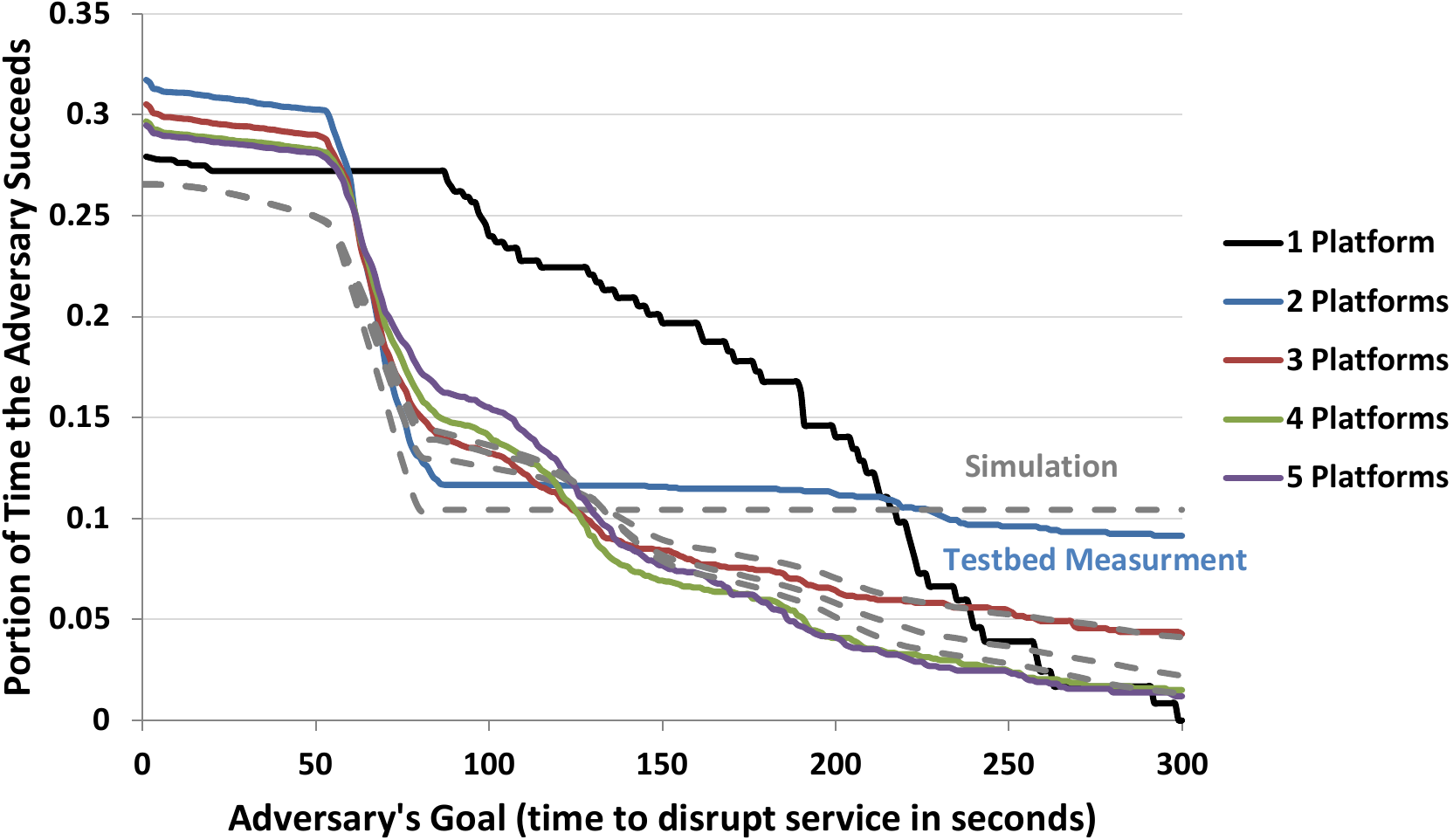}
\end{centering}
\caption{The portion of time that the attacker is in control }
\label{FigureResults1}
\end{figure}

We calculate the value of the metric, which is the percentage of time that the attacker is in control for longer than $T$ and present these results in Fig.~\ref{FigureResults1}.  At the first look, they seem completely counter-intuitive. For small attacker goals ($T$), fewer platforms actually perform better. This is due to the fact that in situations where the attacker wins quickly, more platforms present a larger attack surface. As a result, the attacker wins if she can compromise any of the platforms. In other words,

$\frac{T}{t_{img}}\rightarrow 1$ : Attacker wins iff \textit{any} platform is vulnerable

For the baseline case (1 platform), there is no change in the platform, so the primary effect being observed is the finite duration effect discussed in Section \ref{fd}. The one platform line, in fact, approximates Fig.~\ref{baseline}. 

In addition, notice that the fractional effect (i.e.~downward steps) discussed in Section \ref{fe} can also be observed for two or more platforms.

The value of diversity can only be observed for attacker goals that are large with respect to the inter-migration time ($T\gg t_{mig}$). This is an important parameter when deploying temporal platform diversity systems; the inter-migration time must be selected short enough based on the service requirements of the system. For example, if the system has to survive and provide service within 5 minutes (i.e.~the attacker goal is disrupting service longer than $T = 5$ minutes), the inter-migration time must be $t_{mig}<<5$ min. In other words,

$\frac{T}{t_{img}}\rightarrow N$ : Attacker wins iff \textit{all} platforms are vulnerable

\section{Optimal Control}
\label{GTOC}

While understanding the optimal number of platforms will yield improved performance, the question of \emph{which} platforms to deploy is of critical importance. Recall that in Section \ref{TM} we describe a threat model in which the defender has no knowledge of the vulnerability states of the available platforms. However, even if there exists some vague sense of `platform A is more secure than platform B', it is not always the case that one should deploy platform A. The requirement for adversary persistence results in the system security being achieved through platform diversity.

As an illustration, suppose we have at our disposal three different operating systems -- FreeBSD, Fedora, and CentOS -- and we are able to deploy any two of these in a platform migration system. It would be far better to deploy FreeBSD in combination with either Fedora or CentOS than leveraging the combination of Fedora and CentOS. This is because the latter two OSes are distributions of RedHat Linux and share much of the same kernel code and device drivers. As such, a vulnerability on Fedora is much more likely to exist on CentOS (and vice versa) than on FreeBSD, which is Unix-based but not a Linux distribution. Hence, a single exploit is unlikely to work on both FreeBSD and Fedora/CentOS; this increases the cost for an adversary to compromise the system.

Note that the above illustrative example makes no mention of the security level of the individual platforms. Recall that the goal of the defender is not to globally minimize system vulnerability, but to prevent persistent vulnerability for a duration of length $T$. This is best accomplished by, at each migration time, assuming the current platform is vulnerable and migrating to a platform that is most likely to break the chain of adversary persistence. Intuitively, if the defender knows an adversary has been able to exploit some unknown vulnerability on the current platform, the proper course of action is to migrate to the platform least similar.

Let us formalize this notion by recalling $\mathcal{S}=\{s_{1},\ldots,s_{m+n}\}$ as the set of $m+n$ available platforms and $v(s)$ as a Boolean function specifying whether or not platform $s$ is vulnerable to some available exploit. In general, this function is unknown. Without loss of generality, we discretize time such that an adversary needs to be present for $K=T/t_{mig}$ intervals, where $t_{mig}$ is the specified time between migrations. For simplicity of analysis, we assume that the time required to migrate platforms is negligible. This time is indeed unimportant for our analysis, however it would be if one were to optimize $t_{mig}$, which is an area for future work.

Let us first study the case where $K=2$, for which a first-order Markov chain accurately models the system, then move to the general case. At each interval $k$, an optimal system will migrate to the platform which solves the following:
\begin{equation}
s^k=\arg\min_{j}P\left(v(s_{j})|v(s^{k-1})\right),\label{eq:SingleStep}
\end{equation}
where $s^k\in\mathcal{S}$ is the platform presented during interval $k$. Intuitively, this selects the platform statistically most diverse from the current one. This similarity can be defined in a variety of methods, such as common lines of code or shared modules. If there existed some mapping $f:\mathcal{S}\rightarrow\mathbb{R}^{2}$, Eq.\ (\ref{eq:SingleStep}) would be solved by selecting the point that is furthest away.

Solving the general case for any given $K$, one should assume each of the past $K-1$ platforms have been vulnerable, and therefore exploited. As discussed in Section \ref{AA}, this becomes a Markov chain of order $K-1$, and the optimal platform at each interval $k$ is determined by solving
\begin{equation}
s^k=\arg\min_{j}P\left(v(s_{j})|v(s^{k-1}),v(s^{k-2}),\ldots,v(s^{k-K+1})\right).\label{eq:MultiStep}
\end{equation}
This formulation selects the platform that is jointly most dissimilar from the set of $K-1$ prior platforms. Once again assuming $\exists f:\mathcal{S}\rightarrow X=[x_{1},\ldots,x_{n}]$, where $X\in\mathbb{R}^{2}$, Eq.\ (\ref{eq:MultiStep}) selects the point which maximizes the area of the polygon constructed by joining the set of previous platforms
\begin{equation}
x^k=\arg\max_{j}\textrm{area}(x_{j},x^{k-1},\ldots,x^{k-K+1}).\label{eq:MaxArea}
\end{equation}
We note that Eq.\ (\ref{eq:MultiStep}) can be solved directly, without mapping to a Euclidean space, as there are several methods of computing area between points given just the distances between them. For example, with $K=3$, Heron's formula can be used to compute the area of a triangle.

\subsection{Deterministic Strategy}

It is important to note that Eq.\ (\ref{eq:MultiStep}) results in a periodic scheduling strategy, which can be seen in Eq.\ (\ref{eq:MaxArea}). As each sample point $x_{i}$ is deterministically defined, there will exist a set of $K$ points that form the polygon with the largest area. Regardless of which platform the system is initiated with, the migration pattern will eventually devolve into a periodic rotation across these $K$ platforms. We note that this determinism is acceptable under the adversary model, as the adversary launches exploits against all platforms simultaneously. Hence, randomization offers no benefit to system security in this case.

This relation is important to note, as the typical thought process has been that platform diversity is achieved through randomization. In fact, the opposite is true; randomization and diversity are two
different criterion that are often orthogonal to one another. A strategy that optimizes for randomization would uniformly select from the available platforms, and would almost surely -- as $k\rightarrow\infty$ -- select consecutive platforms that are highly similar and vulnerable to the same exploits. The logic behind this fact was discussed in Section \ref{ABPM}. Contrarily, a strategy that optimizes for diversity returns a deterministic schedule which minimizes the probability of consecutive platforms with similar vulnerabilities, according to some measure of similarity.

\section{Simulation}
\label{Sim}

\begin{table}[t]
\begin{centering}
\begin{tabular}{|r||r@{\extracolsep{0pt}.}l|r@{\extracolsep{0pt}.}l|r@{\extracolsep{0pt}.}l|r@{\extracolsep{0pt}.}l|r@{\extracolsep{0pt}.}l|}
\hline 
 & \multicolumn{2}{c|}{CentOS} & \multicolumn{2}{c|}{Fedora} & \multicolumn{2}{c|}{Debian} & \multicolumn{2}{c|}{Gentoo} & \multicolumn{2}{c|}{FreeBSD}\tabularnewline
\hline 
\hline 
CentOS & 1&0 & 0&6645 & 0&8067 & 0&6973 & 0&0368\tabularnewline
\hline 
Fedora & 0&6645 & 1&0 & 0&5928 & 0&8658 & 0&0324\tabularnewline
\hline 
Debian & 0&8067 & 0&5928 & 1&0 & 0&6202 & 0&0385\tabularnewline
\hline 
Gentoo & 0&6973 & 0&8658 & 0&6202 & 1&0 & 0&0330\tabularnewline
\hline 
FreeBSD & 0&0368 & 0&0324 & 0&0385 & 0&0330 & 1&0\tabularnewline
\hline 
\end{tabular}\caption{Code Similarity Scores, $S(i,j)$, including device drivers \label{tab:Code-Similarity-Scores}}
\par\end{centering}
\end{table}

We now demonstrate the optimality of our presented control strategies through a Monte Carlo simulation, leveraging the pool of five platforms presented in Section \ref{Exper} . We use the Measures of Software Similarity (MOSS)\footnote{http://theory.stanford.edu/$\sim$aiken/moss/} \cite{moss} tool to compute a similarity score between each pair of operating systems. The results are presented in Table \ref{tab:Code-Similarity-Scores}, where each similarity is on a scale of (0-1) where 1.0 implies identical code and 0.0 implies entirely distinct code. The input for each operating system was the kernel code and a set of standard device drivers. As discussed earlier, one should notice that FreeBSD is highly dissimilar to the 4 Linux distributions presented.

During this simulation, we set $K=3$, such that the adversary must be present for 3 consecutive time intervals. This could be viewed, for example, as $T=90$ seconds with $t_{mig}=30$ seconds. Rather than require an extensive set of vulnerability exploits, we define our adversary by randomly selecting one of the five available platforms as vulnerable, $v(s')$. Next, we determine the vulnerability status of each other platform $i$ by performing a Bernoulli trial with a probability of success equal to the similarity score between $s'$ and $s_{i}$. The intuition behind this setup is that the greater the similarity between platform code, the more likely they share a vulnerability. While this is not a direct mapping, the intuition enables a robust analysis, and different measures of similarity could be used.

During each of 500 Monte Carlo (MC) trials lasting 100 time intervals each, we select the vulnerable platforms in the above manner and evaluate three different methods of selecting which platform to migrate to:
\begin{itemize}
\item \emph{Diversity}: Optimizing diversity through Eq.\ (\ref{eq:MultiStep})
\item \emph{Uniform}: Uniform random selection at each interval from the fully available set without immediate repeat
\item \emph{Random 3}: Uniform random selection of $K=3$ platforms prior to the trial, then periodic rotation between them
\end{itemize}
In order to optimize diversity, we define Eq.\ (\ref{eq:MultiStep}) with Heron's formula to compute the area of a triangle, defining the distance between points as $d(i,j)=1-S(i,j)$, where $S(i,j)$ is the similarity score between $s_{i}$ and $s_{j}$ given in Table \ref{tab:Code-Similarity-Scores}.

\begin{figure}
\begin{centering}
\includegraphics[width=0.95\columnwidth]{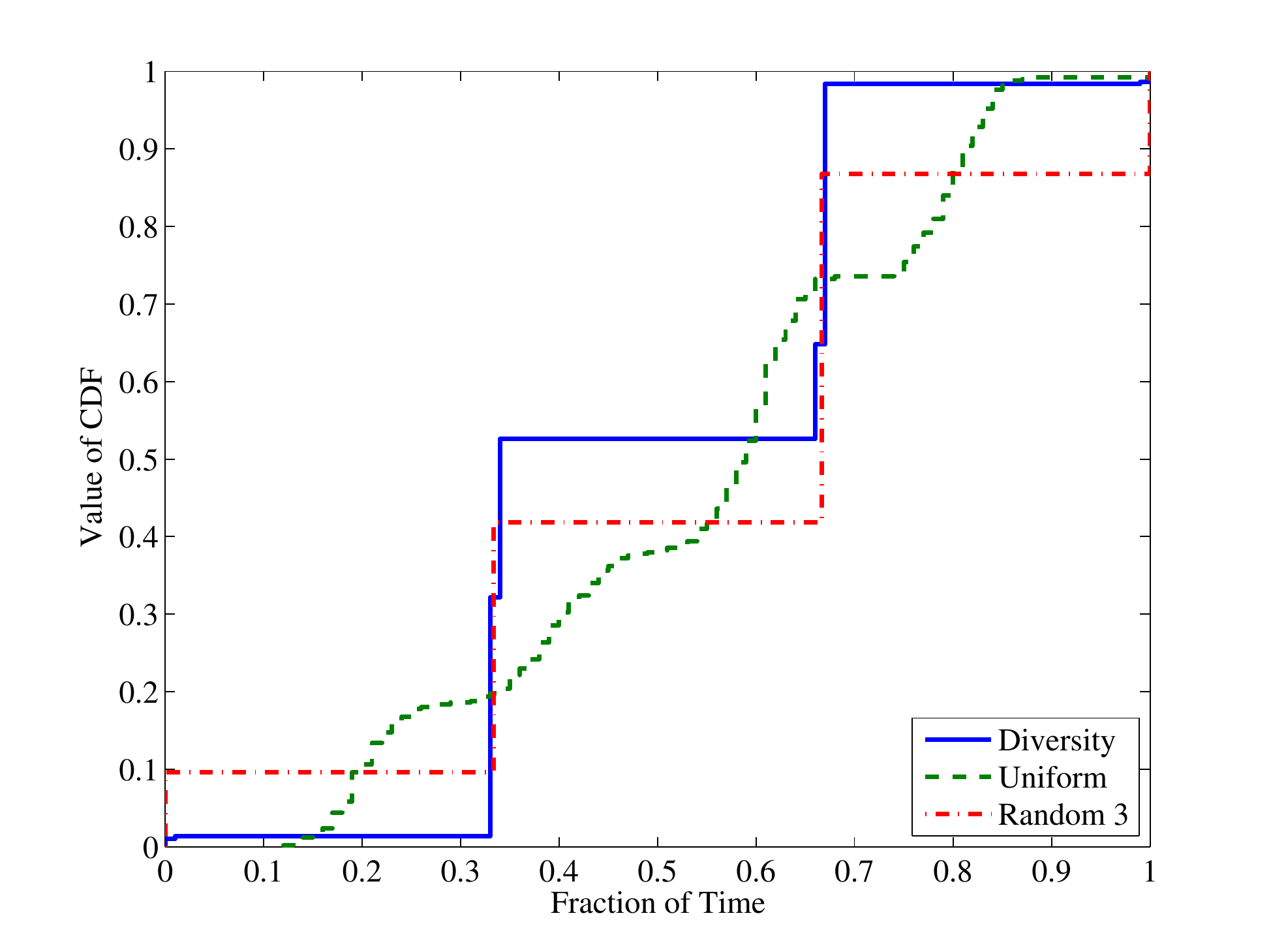}
\par\end{centering}
\caption{Fraction of time presenting a vulnerable system\label{fig:Percentage-Vulnerable}}
\end{figure}

\begin{table}[t]
\begin{centering}
\begin{tabular}{|c|c|c|}
\hline 
Diversity & Uniform & Random 3\tabularnewline
\hline 
\hline 
0.493 & 0.541 & 0.539\tabularnewline
\hline 
\end{tabular}
\par\end{centering}
\caption{Mean vulnerability rate (1-AUC)\label{tab:Mean-vulnerability-rate}}
\end{table}

\begin{figure*}[t]
\centering
\subfloat[Time to first compromise \label{fig:Time-to-first}]{
\includegraphics[width=0.45\textwidth]{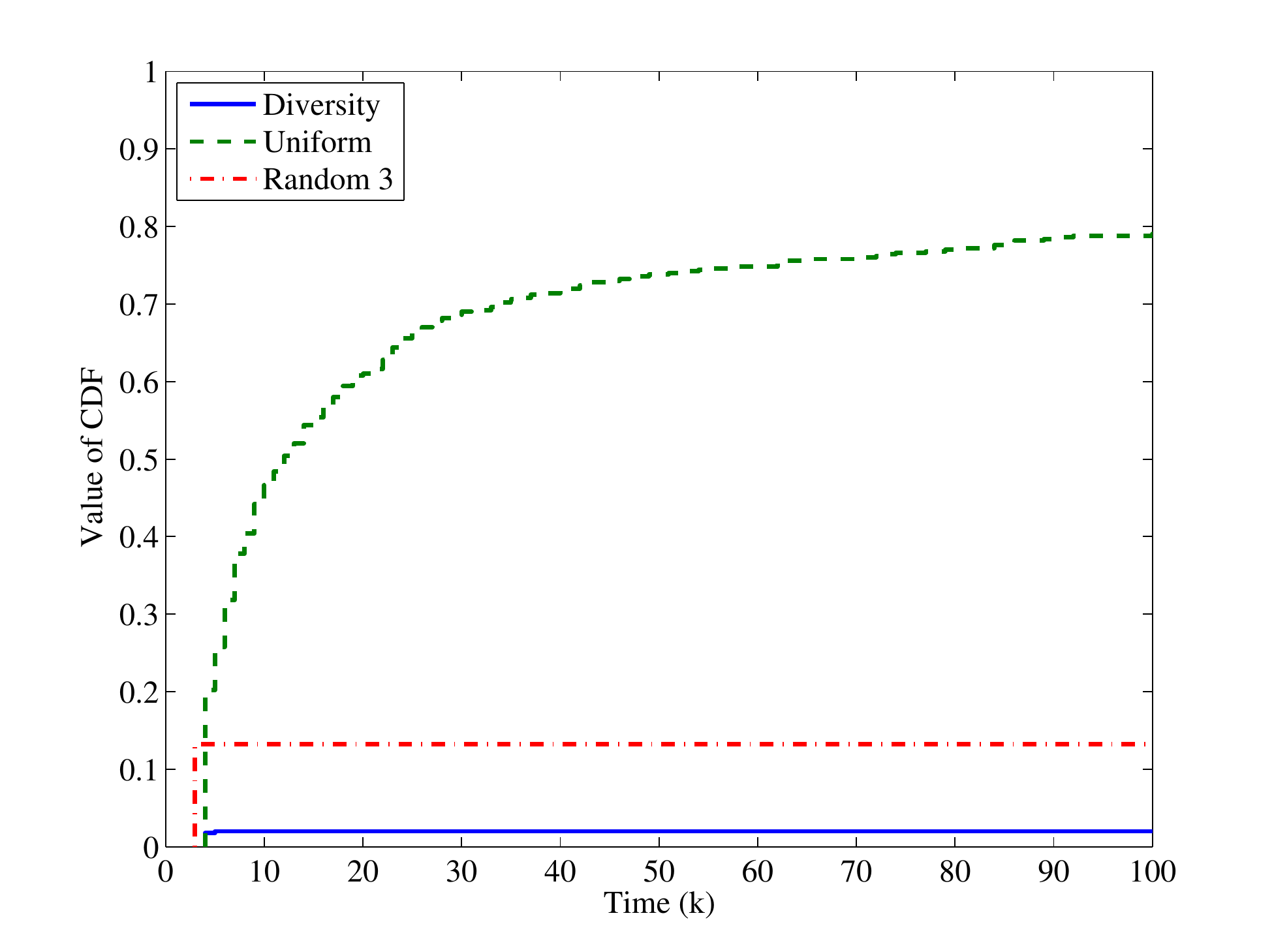}
\par
}
\subfloat[Fraction of time compromised\label{fig:Percentage-Compromised}]{
\includegraphics[width=0.45\textwidth]{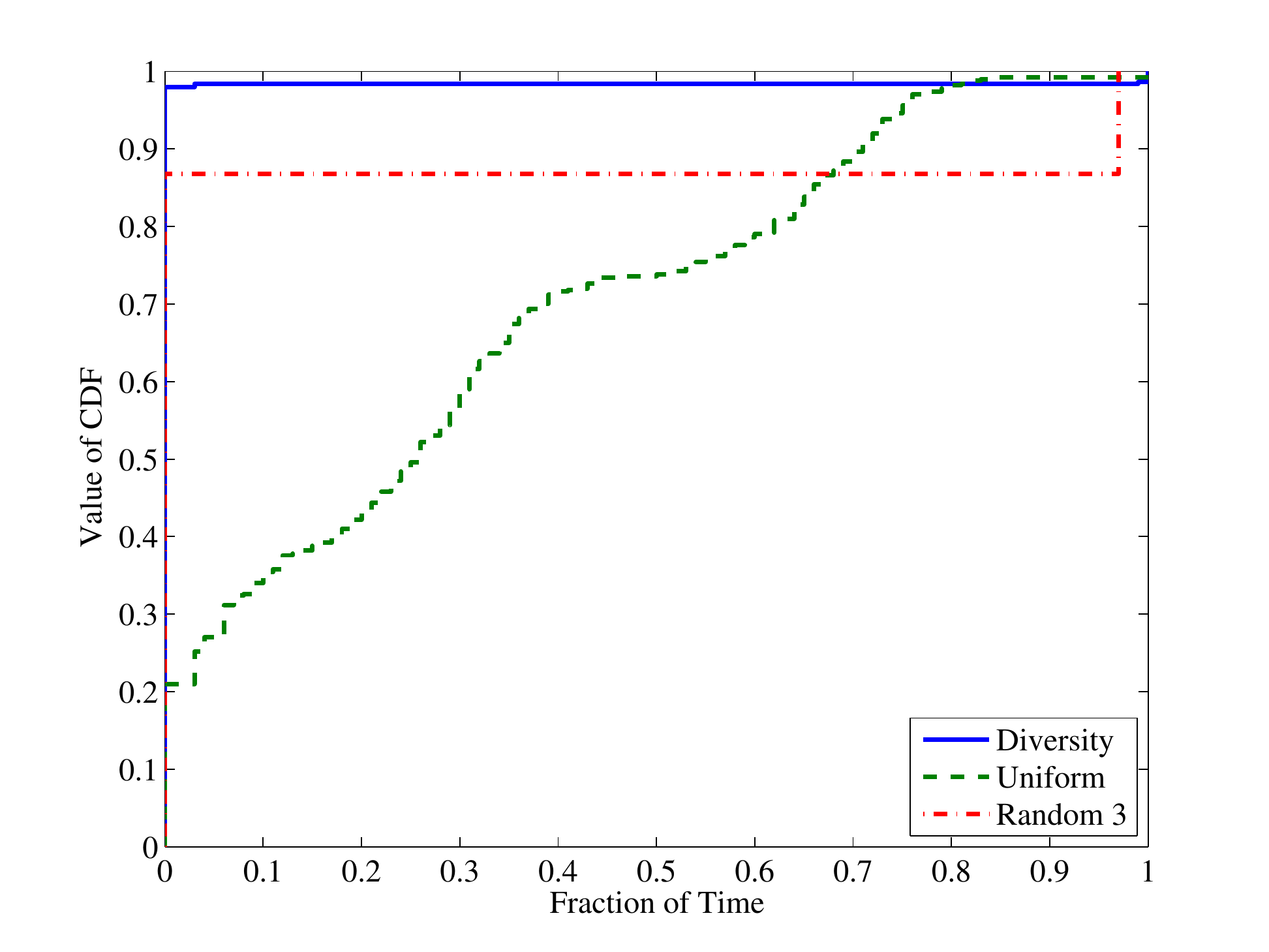}
\par
}
\caption{Evaluation metrics on different migration strategies\label{fig:Evaluation-metrics-Sim}}
\end{figure*}

In Fig.\ \ref{fig:Percentage-Vulnerable} we plot the cumulative distribution function (CDF) of the fraction of time the selected platform was vulnerable, computed across all 500 MC trials. In this plot, a curve trending towards the upper-right corner is preferred, as it demonstrates that a vulnerable system was presented less often. We see that optimizing diversity generally selects less vulnerable systems than either of the other methods; the mean rate of vulnerability for the three methods is shown in Table \ref{tab:Mean-vulnerability-rate}. While diversity does show statistically significant improvement over other strategies, the conclusion from this evaluation would be that none of the strategies are particularly viable. Indeed, in Section \ref{AA} we determined that $K$ was the optimal number of platforms to use, yet its performance is nearly identical to leveraging more. As discussed earlier, however, this is a flawed metric that doesn't represent this attacker model.

We now demonstrate the CDFs of the two metrics which are appropriate for evaluating the given threat models in Fig.\ \ref{fig:Evaluation-metrics-Sim}. First, we study the case for which a single compromise is considered success for the adversary and failure for the defender (e.g.\ `Crash the Satellite'). In Fig.\ \ref{fig:Time-to-first}, we plot the CDF for the time it takes before the system is fully compromised, that is, $K$ vulnerable platforms are presented consecutively. Noting that these results are from the same trials as Fig.\ \ref{fig:Percentage-Vulnerable}, we see a significant difference in performance with the appropriate metric. Optimizing for diversity results in system compromise only 2\% of the time, while uniform selection eventually results in system compromise in 79\% of the trials. We note that because the diversity optimization strategy results in periodic scheduling, if system compromise is going to happen, it does so nearly instantaneously -- either at $k=K$ or shortly thereafter (once 
the 
deterministic solution is reached). This is due to the fact that the system can only be compromised if all of the platforms in the resultant solution are vulnerable; nothing
changes after the solution is reached. The same can be said about the Random 3 strategy, but we note that performance is significantly reduced due to the increased probability of drawing 3 platforms that are highly similar. Contrarily, the Uniform strategy will almost surely result in system compromise as $k\rightarrow\infty,$ if there exists $K$ vulnerable platforms in the fully available set.

Finally, we study the threat model for which an adversary achieves progressive gains the longer they are present on the system (e.g.\ `Exfiltration over a Difficult Channel' and 'Sneak Past a Sensor'). Recall that in these scenarios, the defender loss begins -- or continues -- after the adversary is present for $K$ consecutive intervals. As such, an appropriate evaluation metric is the fraction of time that the system is in a compromised state, which we plot the CDF for in Fig.\ \ref{fig:Percentage-Compromised}. Once again, optimizing diversity yields far superior performance to other methods. In 98\% of the trials, the system was never compromised (note this matches Fig.\ \ref{fig:Time-to-first}), while in that other 2\% it was compromised for the entire duration. Meanwhile, uniform migration results in system compromise on average 26\% of the time, and only in less than 2\% of the trials did it result in less compromise than diversity optimization (13\% of trials when comparing to Random 3).

One can see that when evaluating with the appropriate metrics, the difference in performance becomes significant. This is important to take note of and emphasizes the need to develop a deployment strategy that is optimal towards the specific threat model of interest. Given the threat model requiring adversary persistence, deterministically optimizing for diversity shows superior performance to attempting to achieve security through randomization. We note that while the effects discovered in these simulations match the theoretical results presented in Section \ref{AA} as well as in the experimental results in Section \ref{SS:ExperimentalResults}.

\section{Lessons Learned and Discussions}
\label{LLD}

Our work in analyzing and testing temporal moving target technologies has provided three main lessons.

The first is that the methodology of producing threat derived metrics, as in \cite{lrmetrics}, can be extended to
evaluate base technologies.  In the cases, technologies are designed to address a specific problem or set of problem.
As such security technologies implicitly carry the set of threats which they are intended to address.  These threats can
be used to derive metrics.

While the threats addressed by temporal moving target technologies vary, there remains a certain similarity in the
produced metrics and, perhaps more importantly, in the data needed to evaluate them.  This allows the transformation of
data gathered from one set of, potentially expensive or time consuming, tests in a specific usage scenario to different
usage scenarios.

The second is the surprising result that security, as measured using the threat derived metrics, is not a strictly
increasing as a function of diversity.  Indeed, the minimal diversity required to cover the attackers required duration
provides the optimal solution in the continuous requirement case.  In the aggregate case, the optimal diversity is
determined via a process similar to gerrymandering: that is that the attacker just barely loses in the majority of
sub-selections.

The final lesson is that when we have information concerning diversity between individual platforms, the optimal result
is provided by maximizing platform dissimilarity while minimizing the diversity in the sense of platform count.  Indeed,
even given knowledge of the probability of vulnerability in the various platforms, the most secure ensemble might not be
comprised of the most secure individual members.

\section{Related Work}
\label{RW}

Various platform diversity techniques have been proposed in the literature. As mentioned earlier, The Self-Cleansing Intrusion Tolerance (SCIT) project rotates virtual machines to reduce the exposure time. SCIT-web server \cite{scit-web} and SCIT-DNS \cite{scit-dns} preserve the session information and DNS master file and keys, respectively, but not the internal state of the application. The Resilient Web Service (RWS) Project \cite{rws} uses a virtualization-based web server system that detects intrusions and periodically restores them to a
pristine state. Certain forms of server rotation have been proposed by Blackmon and Nguyen \cite{hafs} and by Rabbat, et al. \cite{cluster} in an attempt to achieve high availability servers. 

High-level forms of temporal platform changes have been proposed by Petkac and Badger \cite{it1} and Min and Choic \cite{it2} to build intrusion tolerant systems although the diversification strategy is not as detailed in these efforts.

Compiler-based multivariant \cite{compilergenerated,reversestack,multivariant,multivariant2,multivariant3} and N-variant systems \cite {nvariant} propose another way of achieving platform diversity.

Holland, et al propose diversifying machine descriptions using a virtualization layer \cite{machinedesc}. A similar approach with more specific diversification strategy based on instruction sets and calling sequences has been proposed by Williams et al \cite{genesis}.

Wong and Lee \cite {cache} use randomization in the processor to combat side-channel attacks on caches. 

On the evaluation side, Manadhata and Wind \cite {attacksurface} propose a formal model for measuring a system's attack surface that can be used to compare different platforms. Evans et al \cite{mteffect} develop models to measure the effectiveness of diversity-based moving target technique. They evaluate the probability of attack success given the time duration of attack probing, construction, and launch cycles and the entropy of randomness in the target system. They evaluate the impact of various attacks on moving target systems including circumvention, deputy, brute force, entropy reduction, probing, and incremental attacks.

Colbaugh and Glass \cite{Colbaugh&Glass:COEX12} use game theory to analyze strategies for deploying moving target defenses against adaptive adversaries, concluding with the result that uniform randomization is optimal. While these results may seem counter to our own, they are not. Their threat model did not require adversary persistence and the defense strategy was designed to minimize predictability. Specifically, the more an adversary is able to observe a defense, they are more likely to adapt to it. In the case of temporal platform migration, this implies that the adversary will expend resources towards the development of exploits specific to the systems observed. While their adversary starts with no capabilities and develops them over time, we instead assume the adversary already has access to a full suite of exploits. One can view these analyses as complimentary, where \cite{Colbaugh&Glass:COEX12} applies to the attacker/defender evolution, while the work we present here applies to the steady state.

\section{Conclusion}
\label{Conc}
In this paper, we have quantitatively studied cyber defenses based on temporal platform diversity. Although the experiments were focused on a migration-based platform diversity system, much of the analyses apply to any such system. Our abstract analysis has studied the major properties in a platform diversity system including the temporal effects, the payoff models, and the window of opportunity using combinatorics and Markov models. Additionally, the impact of finite duration and fractional effects has been studied. The experiments have collected data from a testbed implementation of such a system in which all of the abstract effects can be observed. The game theoretic analysis has studied the impact of preferential platform selection on the effectiveness of such systems and derived an optimal operational strategy. Finally, by simulating the model of the system, we evaluate the impact of different strategies on the effectiveness of temporal platform diversity.

Our results suggest that while platform diversity is useful for mitigating some attacks, it is of critical importance to understand the threat model one aims to defend against. Designing and deploying defensive techniques that have not considered the threat model may result in the illusion of security, when one may actually be increasing their attack surface. Moreover, coupling a platform diversity system with a control mechanism that uses optimal strategies can significantly improve its effectiveness. No one technology will defend against all attacks, so understanding the proper operation and strengths of a platform diversity system enables network defenders ensure that other techniques are deployed for areas of weakness.

The future work in this domain will focus on performing more experiments with such systems, extending the analysis to other platform diversity techniques and other randomization and diversity approaches, and analyzing the second order behavior such as adaptive adversaries who change tactics based on the deployed defenses.

\section{Acknowledgements}
Special thanks to Mark Rabe of MIT Lincoln Laboratory for his help in setting up the experimental test bed.

\newpage

\appendix
\section{Binary Payoff Derivation}
\label{derivation}
Let $L_{v}\left(n\right)$
be the expected length of a sequence beginning with a vulnerable state
that requires $n$ consecutive vulnerable states to fail. Let $L_{I}\left(n\right)$
be the equivalent construct for sequences beginning with an invulnerable
state. Then the mean time until attacker victory is $L(n)=P_{v}\cdot L_{v}\left(n\right)+\left(1-P_{v}\right)\cdot L_{i}\left(n\right)$.
Due to the Markov property we know that $L_{i}(n)=E\left(I^{\left\{ 1\ldots\infty\right\} }\right)+L_{v}(n)$.
Also due to the Markov property, 
\[
\begin{split}
L_{v}(n)=&P_{vv}^{n-1}\cdot E\left(V^{n}\right)+\\
	& \left(1-P_{vv}^{n-1}\right)\cdot\left(E\left(V^{\left\{ 1\ldots n-1\right\} }I^{\left\{ 1\ldots\infty\right\} }\right)+L_{v}\left(n\right)\right)
\end{split}
\]
We can compute the expected lengths of the sequence $I^{\left\{ 1\ldots\infty\right\} }$
as

\[
E\left(I^{\left\{ 1\ldots\infty\right\} }\right)=\frac{\sum_{j=1}^{\infty}j\cdot\left(1-P_{ii}\right)\cdot P_{ii}^{j-1}}{\sum_{j=1}^{\infty}\left(1-P_{ii}\right)\cdot P_{ii}^{j-1}}=\frac{1}{1-P_{ii}}
\]
and of $V^{\left\{ 1\ldots n-1\right\} }$ as $E\left(V^{\left\{ 1\ldots n-1\right\} }\right)=$
\[
\frac{\sum_{i=1}^{n-1}i\cdot\left(1-P_{vv}\right)\cdot P_{vv}^{i-1}}{\sum_{i=1}^{n-1}\left(1-P_{vv}\right)\cdot P_{vv}^{i-1}}=\frac{1-n\cdot P_{vv}^{n-1}+\left(n-1\right)\cdot P_{vv}^{n}}{\left(1-P_{vv}^{n-1}\right)\cdot\left(1-P_{vv}\right)}
\]
 which combines to give the expected time until attacker compromise
{\small
\[
\begin{split}
L(n)=&n+\frac{1-P_{v}}{1-P_{ii}}+\\
	&\left(P_{vv}^{1-n}-1\right)\left(\frac{1-n\cdot P_{vv}^{n-1}+\left(n-1\right)\cdot P_{vv}^{n}}{\left(1-P_{vv}^{n-1}\right)\cdot\left(1-P_{vv}\right)}+\frac{1}{1-P_{ii}}\right)
\end{split}
\]
}

\bibliographystyle{abbrv}

\begin{thebibliography}{10}

\bibitem{HDF5}
{HDF4 Reference Manual}.
\newblock The HDF Group, February 2010.
\newblock {ftp://ftp.hdfgroup.org/HDF/Documentation/
  HDF4.2.5/HDF425\_RefMan.pdf}.

\bibitem{scit3}
D.~Arsenault, A.~Sood, and Y.~Huang.
\newblock Secure, resilient computing clusters: Self-cleansing intrusion
  tolerance with hardware enforced security (scit/hes).
\newblock In {\em Proceedings of the The Second International Conference on
  Availability, Reliability and Security}, ARES '07, pages 343--350,
  Washington, DC, USA, 2007. IEEE Computer Society.

\bibitem{scit}
A.~Bangalore and A.~Sood.
\newblock Securing web servers using self cleansing intrusion tolerance (scit).
\newblock In {\em Dependability, 2009. DEPEND '09. Second International
  Conference on}, pages 60 --65, june 2009.

\bibitem{scit-web}
A.~K. Bangalore and A.~K. Sood.
\newblock Securing web servers using self cleansing intrusion tolerance (scit).
\newblock In {\em Proceedings of the 2009 Second International Conference on
  Dependability}, pages 60--65, 2009.

\bibitem{hafs}
S.~Blackmon and J.~Nguyen.
\newblock High-availability file server with heartbeat.
\newblock {\em System Admin, The Journal for UNIX and Linux Systems
  Administration}, 10(9), 2001.

\bibitem{Colbaugh&Glass:COEX12}
R.~Colbaugh and K.~Glass.
\newblock Predictability-oriented defense against adaptive adversaries.
\newblock In {\em Proceedings of the IEEE Intl. Conference on Systems, Man, and
  Cybernetics}, COEX, pages 2721--2727, 2012.

\bibitem{nvariant}
B.~Cox, D.~Evans, A.~Filipi, J.~Rowanhill, W.~Hu, J.~Davidson, J.~Knight,
  A.~Nguyen-Tuong, and J.~Hiser.
\newblock N-variant systems: a secretless framework for security through
  diversity.
\newblock In {\em Proceedings of the 15th conference on USENIX Security
  Symposium - Volume 15}, USENIX-SS'06, Berkeley, CA, USA, 2006. USENIX
  Association.

\bibitem{mteffect}
D.~Evans, A.~Nguyen-Tuong, and J.~C. Knight.
\newblock Effectiveness of moving target defenses.
\newblock In {\em Moving Target Defense}, pages 29--48. 2011.

\bibitem{machinedesc}
D.~A. Holland, A.~T. Lim, and M.~I. Seltzer.
\newblock An architecture a day keeps the hacker away.
\newblock {\em SIGARCH Comput. Archit. News}, 33(1):34--41, Mar. 2005.

\bibitem{scit2}
Y.~Huang, D.~Arsenault, and A.~Sood.
\newblock Incorruptible system self-cleansing for intrusion tolerance.
\newblock In {\em Performance, Computing, and Communications Conference, 2006.
  IPCCC 2006. 25th IEEE International}, pages 4 pp. --496, april 2006.

\bibitem{rws}
Y.~Huang and A.~Ghosh.
\newblock Automating intrusion response via virtualization for realizing
  uninterruptible web services.
\newblock In {\em Network Computing and Applications, 2009. NCA 2009. Eighth
  IEEE International Symposium on}, pages 114 --117, july 2009.

\bibitem{compilergenerated}
T.~Jackson, B.~Salamat, A.~Homescu, K.~Manivannan, G.~Wagner, A.~Gal,
  S.~Brunthaler, C.~Wimmer, and M.~Franz.
\newblock Compiler-generated software diversity.
\newblock In {\em Moving Target Defense}, pages 77--98. 2011.

\bibitem{multivariant3}
T.~Jackson, B.~Salamat, G.~Wagner, C.~Wimmer, and M.~Franz.
\newblock On the effectiveness of multi-variant program execution for
  vulnerability detection and prevention.
\newblock In {\em Proceedings of the 6th International Workshop on Security
  Measurements and Metrics}, MetriSec '10, pages 7:1--7:8, New York, NY, USA,
  2010. ACM.

\bibitem{openvz}
K.~Kolyshkin.
\newblock Virtualization in linux.
\newblock White paper, OpenVZ, September 2006.

\bibitem{attacksurface}
P.~K. Manadhata and J.~M. Wing.
\newblock A formal model for a system's attack surface.
\newblock In {\em Moving Target Defense}, pages 1--28. 2011.

\bibitem{it2}
B.~J. Min and J.~S. Choi.
\newblock An approach to intrusion tolerance for mission-critical services
  using adaptability and diverse replication.
\newblock {\em Future Gener. Comput. Syst.}, 20(2):303--313, Feb. 2004.

\bibitem{valgrind}
N.~Nethercote and J.~Seward.
\newblock Valgrind: a framework for heavyweight dynamic binary instrumentation.
\newblock In {\em Proceedings of the 2007 ACM SIGPLAN conference on Programming
  language design and implementation}, PLDI '07, pages 89--100, New York, NY,
  USA, 2007. ACM.

\bibitem{nitrd}
F.~Networking, I.~T. Research, and D.~(NITRD).
\newblock {Federal Cybersecurity Game-change R\&D Themes}, 2012.
\newblock {http://cybersecurity.nitrd.gov/page/federal-cybersecurity-1}.

\bibitem{talent}
H.~Okhravi, A.~Comella, E.~Robinson, and J.~Haines.
\newblock Creating a cyber moving target for critical infrastructure
  applications using platform diversity.
\newblock {\em International Journal of Critical Infrastructure Protection},
  5(1):30 -- 39, 2012.

\bibitem{it1}
M.~Petkac and L.~Badger.
\newblock Security agility in response to intrusion detection.
\newblock In {\em in 16th Annual Computer Security Applications Conference
  (ACSAC}, page~11, 2000.

\bibitem{cluster}
R.~Rabbat, T.~McNeal, and T.~Burke.
\newblock A high-availability clustering architecture with data integrity
  guarantees.
\newblock In {\em IEEE International Conference on Cluster Computing}, pages
  178--182, 2001.

\bibitem{cppc}
G.~Rodr\'{\i}guez, M.~J. Mart\'{\i}n, P.~Gonz\'{a}lez, J.~Touri\~{n}o, and
  R.~Doallo.
\newblock Cppc: a compiler-assisted tool for portable checkpointing of
  message-passing applications.
\newblock {\em Concurr. Comput. : Pract. Exper.}, 22(6):749--766, Apr. 2010.

\bibitem{lrmetrics}
{R.P. Lippmann}, {J.F. Riordan}, {T.H. Yu}, and {K.K. Watson}.
\newblock {Continuous Security Metrics for Prevalent Network Threats:
  Introduction and First Four Metrics}.
\newblock Technical report, MIT Lincoln Laboratory, May 2012.

\bibitem{webserver}
A.~Saidane, V.~Nicomette, and Y.~Deswarte.
\newblock The design of a generic intrusion-tolerant architecture for web
  servers.
\newblock {\em Dependable and Secure Computing, IEEE Transactions on}, 6(1):45
  --58, jan.-march 2009.

\bibitem{reversestack}
B.~Salamat, A.~Gal, and M.~Franz.
\newblock Reverse stack execution in a multi-variant execution environment.
\newblock In {\em In Workshop on Compiler and Architectural Techniques for
  Application Reliability and Security}, 2008.

\bibitem{multivariant2}
B.~Salamat, A.~Gal, T.~Jackson, K.~Manivannan, G.~Wagner, and M.~Franz.
\newblock Multi-variant program execution: Using multi-core systems to defuse
  buffer-overflow vulnerabilities.
\newblock In {\em Complex, Intelligent and Software Intensive Systems, 2008.
  CISIS 2008. International Conference on}, pages 843 --848, march 2008.

\bibitem{multivariant}
B.~Salamat, T.~Jackson, G.~Wagner, C.~Wimmer, and M.~Franz.
\newblock Runtime defense against code injection attacks using replicated
  execution.
\newblock {\em Dependable and Secure Computing, IEEE Transactions on}, 8(4):588
  --601, july-aug. 2011.

\bibitem{moss}
S.~Schleimer, D.~S. Wilkerson, and A.~Aiken.
\newblock Winnowing: local algorithms for document fingerprinting.
\newblock In {\em Proceedings of the 2003 ACM SIGMOD international conference
  on Management of data}, SIGMOD '03, pages 76--85, New York, NY, USA, 2003.
  ACM.

\bibitem{strata}
K.~Scott and J.~Davidson.
\newblock {Strata: A Software Dynamic Translation Infrastructure}.
\newblock Technical Report CS-2001-17, 2001.

\bibitem{cache}
Z.~Wang and R.~B. Lee.
\newblock New cache designs for thwarting software cache-based side channel
  attacks.
\newblock In {\em Proceedings of the 34th annual international symposium on
  Computer architecture}, ISCA '07, pages 494--505, New York, NY, USA, 2007.
  ACM.

\bibitem{genesis}
D.~Williams, W.~Hu, J.~W. Davidson, J.~D. Hiser, J.~C. Knight, and
  A.~Nguyen-Tuong.
\newblock Security through diversity: Leveraging virtual machine technology.
\newblock {\em IEEE Security and Privacy}, 7(1):26--33, Jan. 2009.

\bibitem{scit-dns}
A.~S. Yih~Huang, David~Arsenault.
\newblock Incorruptible self-cleansing intrusion tolerance and its application
  to dns security.
\newblock {\em AJournal of Networks}, 1(5):21--30, September/October 2006.

\end{thebibliography}

\end{document}